\documentclass[sigconf]{acmart}

\usepackage{listings} 
\usepackage{xcolor}
\usepackage{mdframed} 
\usepackage{todonotes}
\definecolor{codered}{rgb}{0.6,0,0}
\definecolor{codegreen}{rgb}{0,0.6,0}
\definecolor{codegray}{rgb}{0.5,0.5,0.5}
\definecolor{codepurple}{rgb}{0.58,0,0.82}
\definecolor{backcolour}{rgb}{0.95,0.95,0.92}
\lstdefinestyle{mystyle}{
    backgroundcolor=\color{backcolour},   
    commentstyle=\color{codegreen},
    keywordstyle=\color{magenta},
    numberstyle=\tiny\color{codegray},
    stringstyle=\color{codepurple},
    basicstyle=\ttfamily\footnotesize,
    breakatwhitespace=false,         
    breaklines=true,                 
    captionpos=b,                    
    keepspaces=true,                 
    numbersep=5pt,                  
    showspaces=false,                
    showstringspaces=false,
    showtabs=false,                  
    tabsize=2
}
\usepackage{enumitem}

\AtBeginDocument{%
  \providecommand\BibTeX{{%
    \normalfont B\kern-0.45em{\scshape i\kern-0.25em b}\kern-0.8em\TeX}}}

\copyrightyear{2025}
\acmYear{2025}
\setcopyright{rightsretained}
\acmConference[SIGCSE TS 2025]{Proceedings of the 56th ACM Technical
Symposium on Computer Science Education V. 2}{February 26-March 1,
2025}{Pittsburgh, PA, USA}
\acmBooktitle{Proceedings of the 56th ACM Technical Symposium on Computer
Science Education V. 2 (SIGCSE TS 2025), February 26-March 1, 2025,
Pittsburgh, PA, USA}
\acmDOI{10.1145/3641555.3705214}
\acmISBN{979-8-4007-0532-8/25/02}

\begin{document}




\title[LLM-Generated Analogies]{Lightweight Social Computing Tools for Research Community Building}

\title{Lightweight Social Computing Tools for Building Undergraduate Research Community}

\title{Lightweight Social Computing Tools for Undergraduate Research Community Building}







\author{Noel Chacko}
\affiliation{
 \institution{Temple University}
 \city{Philadelphia}
 \state{PA}
 \country{United States}}
\email{noelchacko@temple.edu}
\orcid{}

\author{Hannah Vy Nguyen}
\affiliation{%
 \institution{Temple University}
 \city{Philadelphia}
 \state{PA}
 \country{United States}}
\email{hannah.nguyen0002@temple.edu}
\orcid{}

\author{Sophie Chen}
\affiliation{%
 \institution{Temple University}
 \city{Philadelphia}
 \state{PA}
 \country{United States}}
\email{sophie.chen@temple.edu}
\orcid{}

\author{Stephen MacNeil}
\affiliation{%
 \institution{Temple University}
 \city{Philadelphia}
 \state{PA}
 \country{US}}
\email{stephen.macneil@temple.edu	}
\orcid{0009-0001-1597-3513}

\renewcommand{\shortauthors}{Noel Chacko, Hannah Vy Nguyen, Sophie Chen, and Stephen MacNeil}

\begin{abstract}


Many barriers exist when new members join a research community, including impostor syndrome. These barriers can be especially challenging for undergraduate students who are new to research. In our work, we explore how the use of social computing tools in the form of spontaneous online social networks (SOSNs) can be used in small research communities to improve sense of belonging, peripheral awareness, and feelings of togetherness within an existing CS research community. Inspired by SOSNs such as BeReal, we integrated a Wizard-of-Oz photo sharing bot into a computing research lab to foster community building among members. Through a small sample of lab members (N = 17) over the course of 2 weeks, we observed an increase in participants' sense of togetherness based on pre and post-study surveys. Our surveys and semi-structured interviews revealed that this approach has the potential to increase awareness of peers' personal lives, increase feelings of community, and reduce feelings of disconnectedness. 

\end{abstract}


\maketitle

\begin{figure}
    \centering
    \includegraphics[width=1\linewidth]{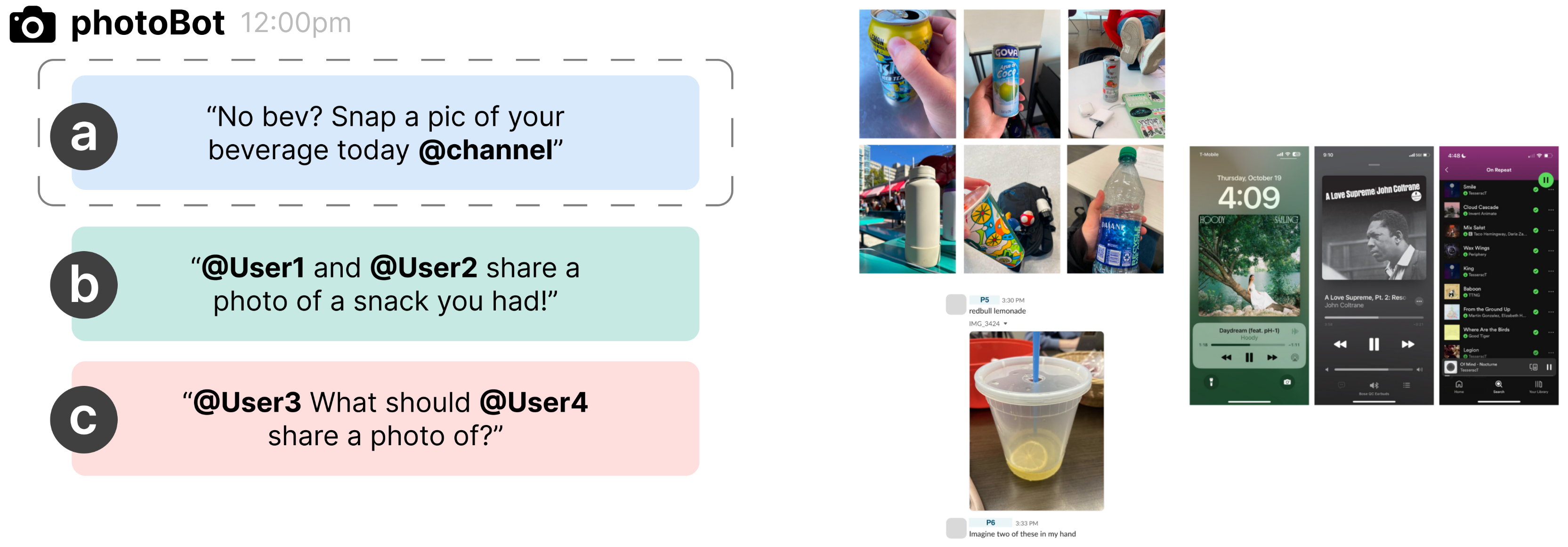}
    \caption{We designed three different prompt types for photoBot. Free Post (A), Selected User(s) (B), and Prompted (C). On the right are examples of photos sent by participants.}
    \label{fig:enter-label}
\end{figure}

\section{Motivation and Background}

Research environments can be intimidating spaces, creating social barriers especially for new members with little to no prior research experience \cite{freeman_coping_2020}. These barriers include impostor syndrome, which can further exacerbate feelings of inadequacy as individuals encounter new ways of working and thinking \cite{freeman_coping_2020}.
Prior research has shown that technology can play a pivotal role in fostering lightweight interactions and increasing awareness among group members\cite{cong_collectivenarrative_2021}, especially via social media platforms. By providing glimpses into peers' daily lives, lab members can recognize shared interests and experiences. Spontaneous online social networks (SOSNs), such as BeReal and Snapchat, enable just-in-time content sharing without the planning associated with regular online social networks (ie: Instagram). We hypothesize that SOSN platforms can assist in mitigating social barriers encountered by new lab members in research and computing communities.

\section{User Study}

\begin{figure*}[ht]
    \centering
    \includegraphics[width=0.92\linewidth]{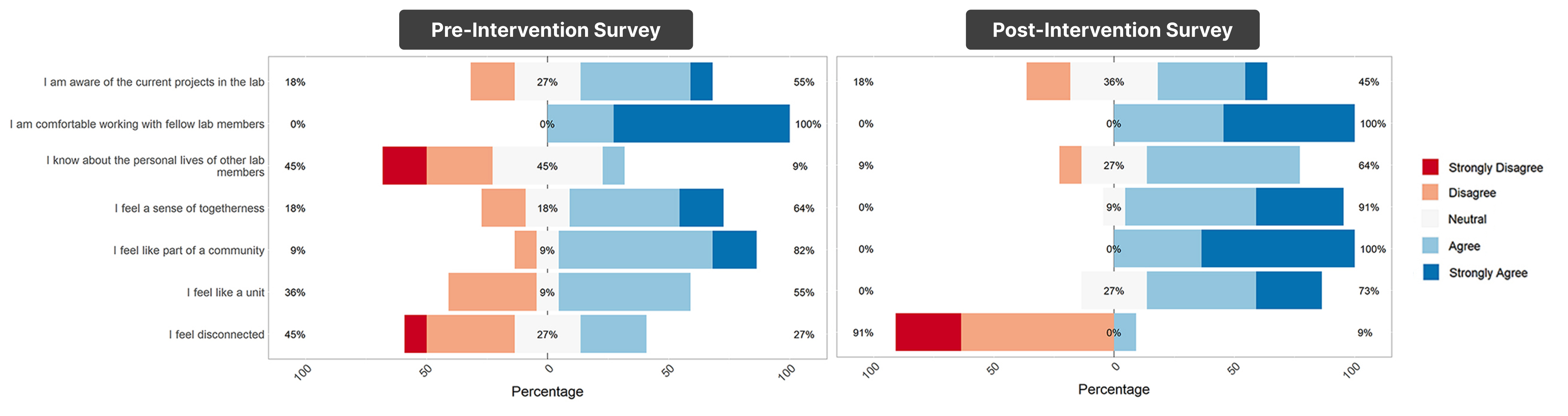}
    \caption{Pre- and post-survey responses show a positive trend toward increased feelings of inclusion.}
    \label{fig:enter-label}
\end{figure*}

To assess how SOSNs may cultivate a deeper sense of togetherness within a computing research lab community, we designed a Wizard-of-Oz `photoBot', a Slack account emulating an intelligent Slackbot sending photo prompts, when in actuality, it was controlled by the research team.
We designed three prompt types.
For the Free Post (\textbf{Fig. 1A}), we prompted any user to post in the channel. For the Selected User(s) (\textbf{Fig. 1B}), we prompted specific users to respond. Finally, in the Prompted condition (\textbf{Fig. 1C}), we asked a user to tag another user(s), creating a prompt for them to respond to. Prompts were scheduled for different times of the day. The photoBot prompted the channel 3 times per day for 14 days.

We recruited 17 undergraduate researchers from our lab's Slack workspace through an announcement in public \textbf{\#general} channel. The lab is located at an R1 university with about 30 members. Participants completed the informed consent and a survey before joining the private \textbf{\#photosharing} channel where the study took place. The research was approved by our Institutional Review Board.


To evaluate participants' engagement with the channel, we used a mixed-methods approach of both semi-structured interviews and surveys. Sample Likert-scale questions from the survey are shown in Figure~\ref{fig:enter-label}. To analyze the interview transcripts, two researchers conducted a collaborative thematic analysis~\cite{braun2019reflecting}, mediating disagreements through discussion. Additionally, we conducted a collaborative content analysis of the shared photos, captions, and reactions.

\section{Results}

In the study, participants made an average of 12.5 posts, with a higher response rate (91\%) when explicitly tagged by the photoBot compared to non-tagged prompts (38\%). Participants shared images, screenshots, and text, often responding more to real-time prompts. The study fostered engagement and conversation, with some participants connecting over shared interests. Although all participants reported getting to know lab members better, only 58\% made new friends, with remote participants feeling more "parasocial" connections. Time of day affected engagement, with participants more likely to post between 11 AM and 3 PM. After the trial ended, 88\% of participants continued responding to prompts, motivated by community-building and learning more about each other. Collaborative thematic analysis of the interviews revealed themes of increased peripheral awareness, participant engagement, and interest in continuing with the \textbf{photosharing} channel. The content analysis revealed that prompts increased conversations between participants about shared hobbies and interests. For example, one of the Free Post prompts was "\textit{Hobbies in action: Share your favorite pastime \textbf{@channel}}". P5 responded with a screenshot of a mobile game they enjoyed, which resulted in a 17-message long conversation with two other participants (P1 and P8).


\section{Discussion}

Given the importance of sense of belonging to academic success~\cite{tinto2012leaving}, computing educators are working to prioritize community-building within classrooms to promote more inclusive learning environments~\cite{latulipe2018evolving, latulipe2015structuring}. Many of these prior works are relatively heavyweight, requiring students to engage in active learning or speed mingling activities. However, our work shows a promising path forward which appears to be effective, lightweight, and low effort. 

Our study highlights that spontaneous photo-sharing increased awareness of peers’ personal lives and fostered connections
Participants reported that the photo-sharing bot made the workspace feel less formal and increased comfort levels, particularly for those who did not physically work in the lab. This sense of belonging extended to hybrid and remote members, who appreciated being able to connect in a more casual, ongoing manner. The lightweight nature of this intervention combined with the role of an external facilitating force such as the photoBot, which drives community interactions, suggests that community sometimes must be pushed by an external factor. Our findings suggest that similar approaches could be easily implemented in other computing education settings without requiring significant resources. By maintaining regular but low-stakes interactions, participants were able to feel more connected over time, which could have a meaningful impact in classrooms, labs, or other hybrid learning environments. 

\section{Limitations and Future Work}

One limitation of our study is the small sample size of 17 participants, which limits the generalizability of our findings. The two-week duration may not capture realistic engagement patterns, particularly during exams or breaks. Pre-existing relationships also influenced engagement. Despite these limitations, this approach shows promise and warrants further investigation, especially in other education environments like introductory CS courses where students may face similar social barriers. Future work could compare our approach to in-person ice-breaker activities.

\bibliographystyle{ACM-Reference-Format}
\bibliography{sample-base}

\balance


\end{document}